\documentclass{sig-alternate-05-2015}

\usepackage{amsmath}
\usepackage{color,graphicx}
\usepackage{url}

\usepackage{caption}
\usepackage{subcaption}
\usepackage[hidelinks=true]{hyperref}

\newsavebox\CBox

\newfont{\mycrnotice}{ptmr8t at 7pt}
\newfont{\myconfname}{ptmri8t at 7pt}
\let\crnotice\mycrnotice%
\let\confname\myconfname%

\usepackage[titlenumbered,ruled]{algorithm2e}

\usepackage{multirow}

\usepackage{tikz}
\usetikzlibrary{matrix,arrows,decorations.pathmorphing}
\usepackage{verbatim}
\usepackage{varwidth}

\usepackage{array}
\newcolumntype{C}[1]{>{\centering\let\newline\\\arraybackslash\hspace{0pt}}m{#1}}
\newcolumntype{L}[1]{>{\let\newline\\\arraybackslash\hspace{0pt}}m{#1}}

\usepackage{algorithmicx}

\usepackage{subcaption}

\begin{document}







%
\conferenceinfo{CIKM Cup 2016}{USA}

\title{Classification and Learning-to-rank Approaches for Cross-Device Matching at CIKM Cup 2016}
%
%
%
%
%

\numberofauthors{1} 
%
\author{
%
%
\alignauthor
Nam Khanh Tran \\
       \affaddr{L3S Research Center - Leibniz Universit{\"a}t Hannover}\\
       \email{ntran@l3s.de}
}

\maketitle
\begin{abstract}
In this paper, we propose two methods for tackling the problem of 
cross-device matching for online advertising at CIKM Cup 2016. The first method considers the matching problem as a binary classification task and solve it by utilizing ensemble learning techniques. The second method defines the matching problem as a ranking task and effectively solve it with using learning-to-rank algorithms. The results show that the proposed methods obtain promising results, in which the ranking-based method outperforms the classification-based method for the task. 

\end{abstract}


%
%

%
%

%
%



\section{Introduction}

Online advertising is to help companies market their products and services to the right audiences of online users. In doing so, advertising companies have to collect a lot of user generated data such as browsing logs and ad clicks, perform sophisticated user profiling, and compute the similarity of ads to user profiles.

However, as the number and variety of different devices increases, the online user activity becomes highly fragmented. People tend to use different devices for different purposes, for example doing work on laptops, reading documents on tablets. The same user is usually viewed independently on different devices. Moreover, even the same device could be shared by many users, e.g. both kids and parents sharing a computer at home. Therefore, building accurate user identity becomes a very difficult and important problem for advertising companies.
 

The CIKM Cup 2016 Track 1: Cross-Device Entity Linking Challenge hosted by CodeLab from August 5th to October 5th is for tackling this problem. The competition attracted 155 registered participants, in which top-5 participants have made over 800 submissions.




Some works have been investigated to solve similar problems. \cite{Lee:2014,Bannan:2014} studied on how to promote advertisement when dealing with the cross-device problem. \cite{Liben-Nowell:2003,Lichtenwalter:2010} focused on the problem of link prediction on a (social) graph. \cite{Jeremy:2015,Landry:2015} are closest to our approaches. While \cite{Landry:2015} proposed a multi-layer classification method, \cite{Jeremy:2015} presented a learning-to-rank method for cross-device connection identification. 

In this paper, we describe our approaches for tackling the cross-device matching problem. In the first method, we formulate it as a binary classification task: Given a pair of userIDs, classify the pair into two groups $\{1,0\}$ which indicates that the userIDs associate with the same user or different users, respectively. In the second method, the problem is defined as a ranking task: Given an userID, rank other userIDs according to their possibilities of referring to the same user.


\section{Dataset}
\label{sec:dataset}
The organizers provided a collection of browsing logs and associated meta-data which are fully anonymized to allay privacy concerns and protect business sensitive information. A browsing log contains a list of events (facts) for a specific userID. Each event is mapped to a hashed URL and a hashed HTML title.\footnote{\scriptsize{The detailed description of each file can be seen on the competition website https://competitions.codalab.org/competitions/11171}}
In addition, a set of matching userIDs is also provided for training any supervised model. The goal of the challenge is to predict matching pairs of testing userIDs, in which the userIDs in the training and testing set do not overlap and about 0.5\% ``noise" userIDs are added to the testing set.
Table \ref{tab:dataset_statistics} presents some statistics of the provided dataset.

\begin{table}
	\centering
	\begin{tabular}{|l | c|	}
	\hline
	Total users & 339,405 \\
	\hline
	Number of users in testing set & 98,255 \\
	\hline
	Number of matching pairs in training set & 506,136 \\
	\hline
	Number of matching pairs in testing set & 215,307 \\
	\hline
	Number of unique tokens in titles & 8,485,859 \\
	\hline
	Number of unique tokens in URLs & 27,398,114 \\
	\hline
	\end{tabular}
	\caption{Dataset statistics}
	\label{tab:dataset_statistics}
\end{table}

\begin{figure}[tb]
	\centering
	\includegraphics[scale=0.35]{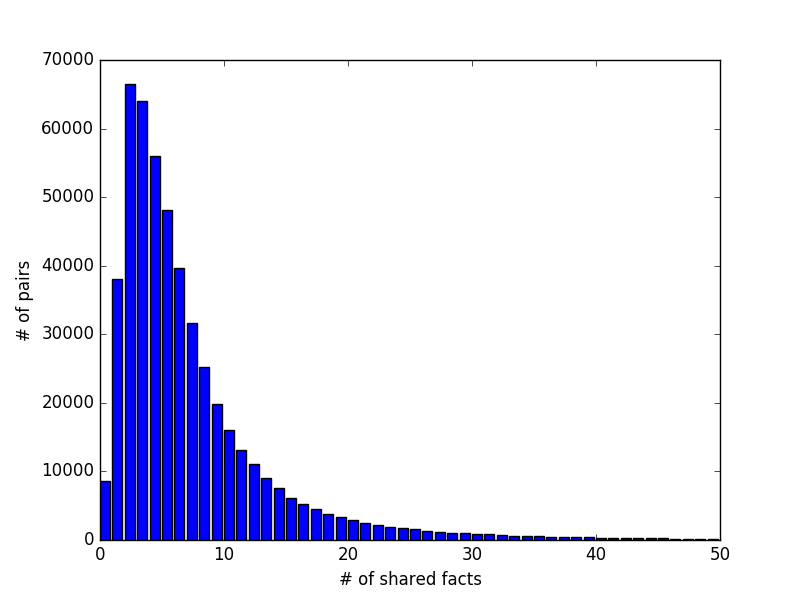}
	\caption{Distribution of number of pairs given the number of shared facts in the training set}	
	\label{fig:train_share_fact}
\end{figure}

\begin{figure}[tb]
	\centering
	\includegraphics[scale=0.35]{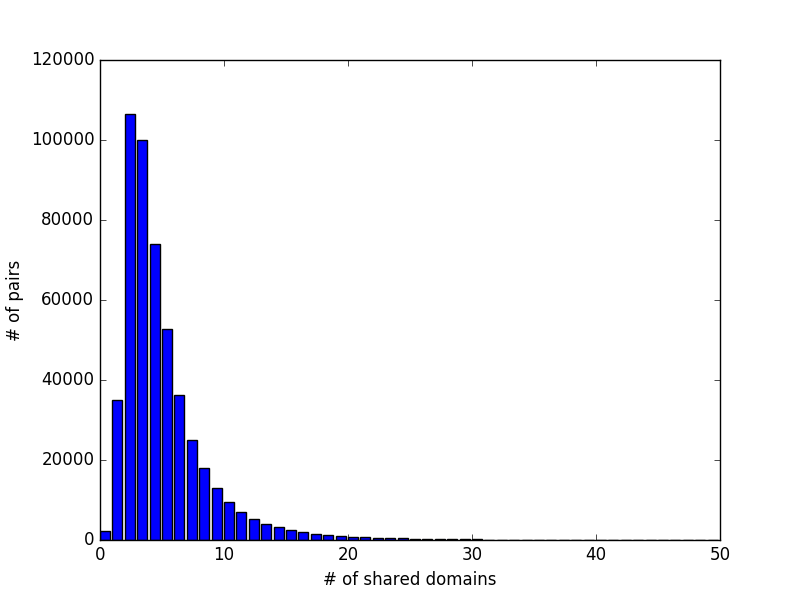}
	\caption{Distribution of number of pairs given the number of shared domains in the training set}	
	\label{fig:train_share_domain}
\end{figure}

Figure \ref{fig:train_share_fact} and Figure \ref{fig:train_share_domain} show the distribution of number of pairs given the number of facts and domains shared between the userID pairs in the training set, respectively. There is only about 1.5\% of pairs does not share any facts, and about 0.4\% of pairs does not access to same domains. These distributions give us a clue that even on different devices users tend to make same actions (facts) or access to the same domains. 

Based on this observation, we propose a negative sampling method to create negative examples for our supervised learning methods as well as generate userID pairs for prediction from the testing set.


\section{Building a training and predicting set}


\subsection{Negative examples}
In the training set, 506,136 matching pairs of userIDs are provided without any negative pairs, thus we need to generate a set of negative pairs for supervised learning algorithms. 

In a very simple way, for each userID $u$ in the training set, we can randomly select an userID $v$ which are not paired with $u$ as negative training samples, which give us a large number of negative pairs (approximately 40B). Consequently, it requires a lot of computation and not feasible for learning algorithms. Therefore, in this work we propose a simple yet effective method for creating negative samples based on the observation discussed in Section \ref{sec:dataset}.

First, for each userID $u$, we create a list of related userIDs $L(u)$ in which each userID in $L(u)$ shares one or more facts/domains with $u$. Then, for each matching pair of userIDs $(u_1,u_2)$ in the training set, we create six different negative pairs, three for the userID $u_1$ and three for the userID $u_2$. The userIDs in the negative pairs are randomly selected from  $L(u_1)$ and $L(u_2)$, respectively; if $L(u_i)$ is empty, they are randomly chosen from the userIDs in the training set. To this end, we obtained a total of 3,5 millions training pairs for our learning algorithms.




\subsection{Prediction set}
The straightforward method for generating prediction set is to pair each userID with all other userIDs in the testing set. However, the number of userIDs in the testing set are large, approximately 98 thousands, consequently we can end up with billions of possible pairs.
Hence, to improve efficiency and reduce computation cost, we propose a method for generating a set of userID pairs for prediction. In practice, we observe that the method can cover almost pairs of userIDs need to predict.

As discussed in Section \ref{sec:dataset}, users tend to create the same events/facts or access to the same domains across different devices such as watching videos on Youtube, accessing to social platforms Facebook/Twitter, etc. Therefore, we restrict our prediction set to only approximately 56 millions of pairs whose userIDs have one or more facts/domains in common.





\section{Approaches}
In this section, we describe two approaches for tackling the cross-device matching problem. In the first approach, we cast the problem to a binary classification task and present an ensemble learning technique to generate the prediction. In the second approach, we formulate it as a learning to rank task, and present a pairwise approach and a graph-based algorithm for generating the prediction. Before describing two approaches in details, we start by presenting our learning features.
\subsection{Learning features}
For each pair of userIDs, we first extract a set of features based on the provided browsing logs and associated meta-data: \textit{fact-based, domain-based, title-based, time-based} and \textit{hybrid} features.
\subsubsection{Fact-based features}
The first type of feature is generated based on the fact/event information. Each user is considered as a document and the facts that she/he made are considered as words of the document. Then, jaccard, tf-idf similarities are computed as learning features.

In addition, we extract topic distributions of documents (users) by applying the topic modeling method i.e. Latent Dirichlet Allocation (LDA)\footnote{\scriptsize{http://mallet.cs.umass.edu/topics.php}}. Each user is now represented by a distribution of topics, then some similarity measures including Hellinger, cosine, jaccard are calculated as additional features. 

Furthermore, we also employ doc2vec\footnote{\scriptsize{https://radimrehurek.com/gensim/models/doc2vec.html}} to represent each user as a vector of 100 dimensions and then calculate similarity measures as learning features.

 
\subsubsection{Domain-based and title-based features}
The domain-based and title-based features are extracted in a similar way as the fact-based features. Each user is considered as a document, however the words of the document are now represented by domains and titles instead of facts.

\subsubsection{Time-based features}
In the same spirit, we represent each user as a sequence of days which he/she had activities, and then calculate jaccard, rf-idf features.

Additionally, we assume that users tend to use different devices at different time. For example, they might use desktop/laptop in the working time such as from 9am to 12am, and mobile phone during the lunch time and their tablets in the evening. Therefore, we extract additional time-based features by calculating the distance between the time (in hour) of userIDs.

\subsubsection{Hybrid features}
This kind of feature takes time information and other meta-data into account, i.e. time with fact, time with domain, for example, extracting time-based features for only shared facts/domains or computing fact-based features for the facts in the same day.

In total, we generate 75 different learning features for our learning algorithms.

\subsection{Prediction as classification}
\begin{figure}[tb]
	\centering
	\includegraphics[trim={0.5cm 4cm 3.5cm 0.5cm},clip=true, scale=0.35]{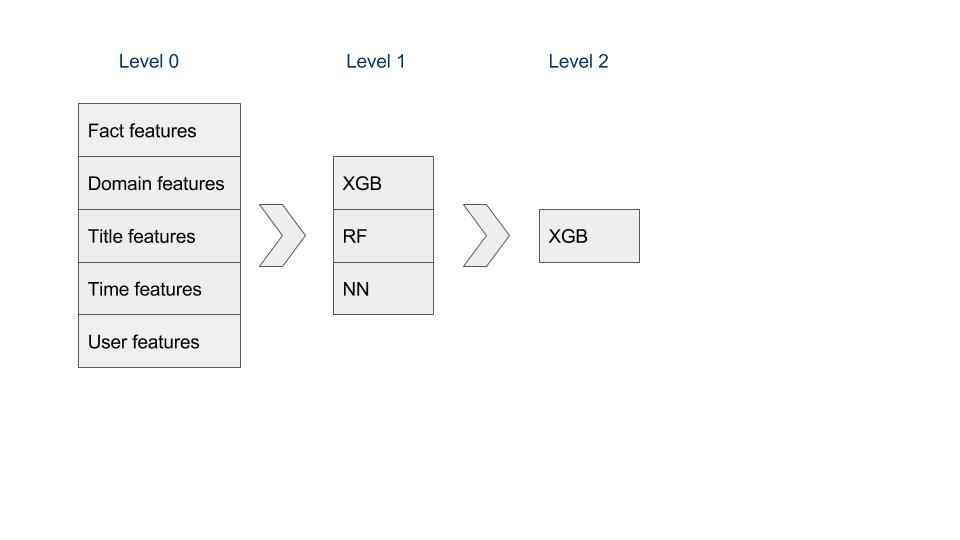}
	\caption{Ensembling learning technique for cross-device entity classification}
	\label{fig:ensemble approach}
\end{figure}
\textbf{Problem} Given a pair of userIDs from the prediction set, classify it into two groups $\{1,0\}$ indicating that they are associated with the same user or different users. The final prediction is generated based on the classification scores of each pair in the prediction set.

\vspace{0.3cm}
\textbf{Method} Figure \ref{fig:ensemble approach} describes an overview of our ensembling learning technique for tackling the cross-device entity matching problem. We start with 75 features extracted for each training pair of userIDs in the Layer 0. These features are used as the inputs for classification algorithms in Layer 1, whose outputs are used as the inputs for the algorithm in Layer 2.

The training examples in Layer 0 are first split into learning and validation sets in which 2/3 examples are used for learning, and 1/3 for validation. The parameters of the algorithms in Layer 1 are learned using the $K$ folds strategy on the learning set and the best configurations are evaluated using the examples in the validation set. 

The classification scores from the algorithms in the Layer 1 are considered as the meta-features for the algorithm in Layer 2. The parameters are again tuned using the examples in the validation set with the $K$ folds method. The best parameter values are then used to calculate prediction scores for the pairs in the prediction set and then generate the top-$N$ pairs for the submission.

In Layer 1, we explore several supervised learning algorithms including Neural Network (NN), Extreme Gradient Bossting (xgboost) and Random Forest, and in Layer 2, we make use of xgboost to compute the final prediction scores.

\subsection{Prediction as ranking}
\textbf{Problem} Given an userID $u$, rank userID $v$ based on the probability of referring to the same user.
In this work, we use LambdaRank, the state-of-the-art learning-to-rank algorithm, to tackle the ranking task. To obtain the final prediction, we propose a simple yet effective graph-based algorithm using the ranking scores obtained from LambdaRank.

\vspace{0.3cm}
\textbf{Method} 
In a pairwise learning-to-rank algorithm \cite{Burges:2010}, we accumulate cost only between pairs of items in the same list. The cost $L_{ij}$ associated with the assigning scores $s_i$ and $s_j$ to examples $i, j$ respectively is:
\[
	L_{ij} = \text{log}(1 + \text{exp}(-f_{ij}(s_i - s_j)))
\]
where $f_{ij}=1$ when $y_i > y_j$, $f_{ij}=-1$ when $y_i<y_j$ and $f_{ij}=0$ when $y_i=y_j$.

Minimizing the cost on the training data, and applying to validation/testing set gives us a ranking of userIDs (documents) given an userID (query) with a associated scores. 

In order to use learning-to-rank algorithms, we need to create datasets with \textit{query} and \textit{documents} information. To obtain this purpose, we treat each userID $u$ in the training/prediction set as query, and other userIDs which are associated with $u$ as documents. Then, we employ LambdaRank to learn the ranking score of the documents (userIDs) for a given query (userID).

For each userID $u_i$, the learning-to-rank algorithm returns a set of weighted userIDs $S(u_i)=\{(v^i_1,w^i_1),...,(v^i_p,w^i_p)\}$ where $v^i_j$ is a possible associated userID of $u_i$ and $w^i_j$ is the association weight. The task now is to generate the most probable matching pairs from this ranking. To tackle this task, we propose two different methods.

\vspace{0.2cm}
\textbf{Rank 1} Sort the pairs of userIDs $(u,v)$ based on its ranking score, and take top-$N$ pairs with the highest scores.

\vspace{0.2cm}
\textbf{Rank 2} The \textit{Rank 1} method is not optimal in the sense that a userID can be associated with zero or hundreds of userIDs. Noted that every userID in the test set is connected with some other userIDs (\textit{constraint 1}). In addition, if we represent each userID as a node, and two userIDs in a pair forms an edge, we obtain a userID graph. The userIDs which refer to the same user form a connected component in the graph.
\begin{figure}[tb]
	\centering
	\includegraphics[scale=0.35]{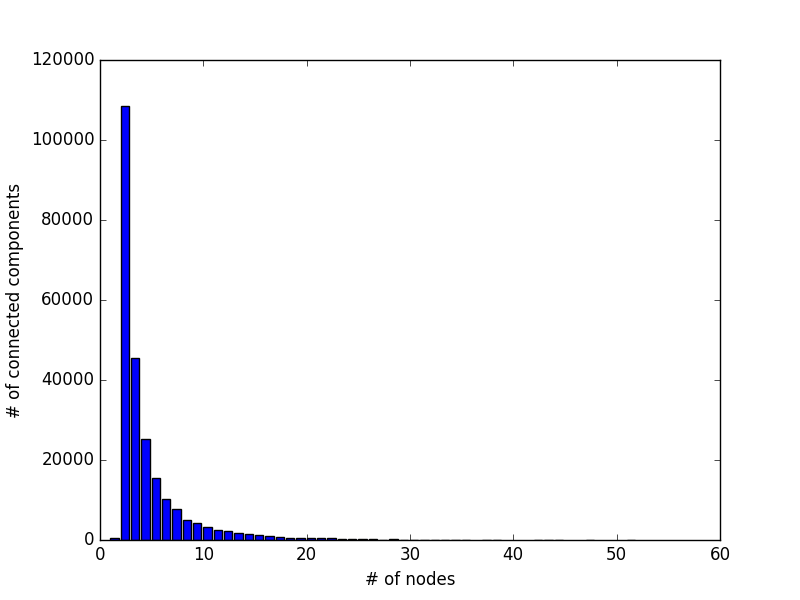}
	\caption{Distribution of the number of connected components over nodes in the training userID graph}
	\label{fig:training graph}
\end{figure} 

Figure \ref{fig:training graph} shows the distribution of the number of connected components given the number of nodes in the training graph. Here, we also want the predictions to follow the same distribution as in the training graph (\textit{constraint 2}). Therefore, we propose a simple yet effective algorithm as shown in Algorithm \ref{alg:greedy} to generate predictions satisfying the constraints.

\begin{algorithm}[t]
  \DontPrintSemicolon
  \SetKwInOut{Input}{Input}\SetKwInOut{Output}{Output}
  \Input{\textnormal{a list of triples} $\{u,v,w\}$, \textnormal{number of pairs to return $N$}}
  \Output{\textnormal{top-}$N$ \textnormal{most probable pairs.}}

\textnormal{ }\\
Sort: $L \leftarrow \{u,v,w\}$ according to weight $w$ \\
P = $list()$ // most probable pairs \\
F = $map()$ // number of associated nodes \\
\For{k in $\{1,2,51\}$}
	{
		\For {(u,v,w) in L}
		{
			\If {F(u) < k and F(v) < k}
			{
				add(u,v) $\rightarrow$ P \\
				increase F(u) \\
				increase F(v) \\
			}
		}
	}

Return top-N pairs in P
\caption{Greedy algorithm for generating final predictions}
\label{alg:greedy}
\end{algorithm}



\subsection{Implementation}
The scripts to generate negative examples, prediction pairs and extract learning features as described in Section 3 are implemented using python\footnote{\scriptsize{https://github.com/namkhanhtran/cikm-cup-2016-cross-device}}. For learning algorithms, we make use of several libraries including scikit-learn\footnote{\scriptsize{http://scikit-learn.org}}, keras\footnote{\scriptsize{https://keras.io/}}, and xgboost\cite{Chen:2016}\footnote{\scriptsize{https://github.com/dmlc/xgboost}}.

\section{Result}
Table \ref{tab:validation} presents the results of classification methods on our validation set. In the validation set, there is 167,055 positive examples, thus the results reported in Table \ref{tab:validation} are computed at top 167,055. We do not care about tuning the number of prediction pair at this point. It can be seen that xgboost obtains the highest scores, while neural network and random forest achieve slightly lower results. Table \ref{tab:validation} also shows that the stacking method outperforms all isolated classification algorithms in the Layer 1, which indicates the effectiveness of stacking learning technique for the classification task.
\begin{table}[t]
	\centering
	\begin{tabular}{|l|c|}
		\hline
		Method & Precision \\
		\hline
		\multicolumn{2}{|l|}{\textbf{Layer 1}} \\
		\hline
		xgboost & 0.927\\
		neural network & 0.911 \\
		random forest & 0.904 \\
		\hline
		\multicolumn{2}{|l|}{\textbf{Layer 2}} \\
		\hline
		xgboost & 0.932 \\
		\hline
	\end{tabular}
	\caption{Performance of classification methods on our validation set}
	\label{tab:validation}
\end{table}

\begin{table}[t]
	\centering
	\begin{tabular}{|l|c|c|}
	\hline
	Method & Precision & Recall \\
	\hline
	\multicolumn{3}{|l|}{\textbf{Layer 1}} \\
	\hline
	xgboost & 0.208 & 0.416 \\
	neural network & 0.195 & 0.390 \\
	random forest & 0.172 & 0.343\\
	\hline
	\multicolumn{3}{|l|}{\textbf{Layer 2}} \\
	\hline
	xgboost & 0.209 & 0.417 \\
	\hline
	\end{tabular}
	\caption{Performance of classification methods on the validation leaderboard (top 215,307 pairs)}
	\label{tab:leaderboard}
\end{table}

Table \ref{tab:leaderboard} shows the results of the classification methods on the provided validation set. It shows a similar pattern to the results on our validation set, in which the stacking method outperforms all individual learning algorithms. However, the performance difference is not large as it is shown on our validation set. 

We develop the learning-to-ranking approach when the validation phase was closed, consequently we can tune our parameters only on our validation set. Therefore, we can only report the results during the testing phase.

\begin{table}[h]
	\centering
	\begin{tabular}{|l|c|c|c|}
	\hline
	Method & F1 & Precision & Recall \\
	\hline
	Rank1 & 0.42038 &	0.39875 &	0.44449 \\
	Rank2 & 0.41669 &	0.39444 &	0.44160 \\
	Rank3 & 0.41370 &	0.40042 &	0.42790 \\
	Rank4 & 0.40168 &	0.36591 &	0.44520 \\
	\hline
	Ours (clf) & 0.29229 & 0.21921 & 0.43843 \\
	Ours (rank$_1$) & 0.29680 & 0.22260 & 0.44520 \\ 
	Ours (rank$_2$) & 0.32761 & 0.24570 & 0.49142 \\
	Ours (rank$_2$-tuned) & 0.36110 &	0.33227 &	0.39540 \\
	\hline
	\end{tabular}
	\caption{CIKM Cup 2016 testing leaderboard}
	\label{tab:final}
\end{table}
Table \ref{tab:final} shows the results of our methods among top-5 participants, in which the learning-to-rank method outperforms the classification methods and the greedy algorithm $rank_2$ obtains better results than the $rank_1$ algorithm.

In addition, as we can submit any number of prediction pairs, tuning this number can boost the rank on the testing leaderboard, the $rank_2$-tuned algorithm.

\section{Conclusions}
\label{sec:Conclusions}
We have presented two different approaches for tackling the cross-device matching problem. In the classification-based approach, the ensemble learning technique shows improvements over the isolated classification algorithms. The ranking-based method with a proposed greedy algorithm for generating the predictions outperforms the classification-based method. Furthermore, we also show that tuning the number of prediction pairs can increase the $F1$ scores and obtain a higher rank in the final ranking.



%
\bibliographystyle{abbrv}
\bibliography{ntran}  
%
%

\end{document}